pdfoutput=1
\def\year{2021}\relax
\documentclass[letterpaper]{article} 
\usepackage{aaai21}  
\usepackage{times}  
\usepackage{helvet} 
\usepackage{courier}  
\usepackage[hyphens]{url}  
\usepackage{graphicx} 
\usepackage{natbib}  
\usepackage{caption} 
\usepackage[misc]{ifsym}

\usepackage{subcaption}
\usepackage[ruled,vlined]{algorithm2e}

\usepackage[mathlines]{lineno}

\usepackage{xr}

\makeatletter
\newcommand*{\addFileDependency}[1]{
  \typeout{(#1)}
  \@addtofilelist{#1}
  \IfFileExists{#1}{}{\typeout{No file #1.}}
}
\makeatother
\newcommand*{\myexternaldocument}[1]{%
    \externaldocument{#1}%
    \addFileDependency{#1.tex}%
    \addFileDependency{#1.aux}%
}
\myexternaldocument{appendix}

\title{Explainability Matters:  Backdoor Attacks on Medical Imaging} 

\author {
        Munachiso Nwadike,\textsuperscript{\rm*1}
        Takumi Miyawaki,\textsuperscript{\rm*1}
        Esha Sarkar,\textsuperscript{\rm 2} \\
        Michail Maniatakos,\textsuperscript{\rm 1}
        Farah Shamout\textsuperscript{\rm 1 $\dagger$}  \\
}
\affiliations {
    \textsuperscript{\rm 1} NYU Abu Dhabi, UAE \\
    \textsuperscript{\rm 2} NYU Tandon School of Engineering, USA \\
    \textsuperscript{\rm *} Equal Contributions\\
    $\dagger$ \texttt{fs999@nyu.edu}
}

\begin{document}

\maketitle

\begin{abstract} 
\begin{quote}
Deep neural networks have been shown to be vulnerable to backdoor attacks, which could be easily introduced to the training set prior to model training. Recent work has focused on investigating backdoor attacks on natural images or toy datasets. Consequently, the exact impact of backdoors is not yet fully understood in complex real-world applications, such as in medical imaging where misdiagnosis can be very costly. In this paper, we explore the impact of backdoor attacks on a multi-label disease classification task using chest radiography, with the assumption that the attacker can manipulate the training dataset to execute the attack. Extensive evaluation of a state-of-the-art architecture demonstrates that by introducing images with few-pixel perturbations into the training set, an attacker can execute the backdoor successfully without having to be involved with the training procedure. A simple 3$\times$3 pixel trigger can achieve up to 1.00 Area Under the Receiver Operating Characteristic (AUROC) curve on the set of infected images. In the set of clean images, the backdoored neural network could still achieve up to 0.85 AUROC, highlighting the stealthiness of the attack. As the use of deep learning based diagnostic systems proliferates in clinical practice, we also show how explainability is indispensable in this context, as it can identify spatially localized backdoors in inference time.

\end{quote}
\end{abstract}

\section{Introduction}

\noindent In recent years, great research has gone into the use of deep learning models in computer-aided diagnostic (CAD) systems. These models have shown promising diagnostic results in many clinical domains, such as fundoscopy \cite{asiri2019deep}, dermatology \cite{liu2020deep,esteva2017dermatologist}, and pulmonary disease diagnosis \cite{shoeibi2020automated,rajpurkar2017chexnet,gabruseva2020deep}. Due to shortages in radiologists worldwide and increased burnout \cite{wuni2020opportunities,ali2015diagnostic, kumamaru2018global,zha2018prevalence,rimmer2017radiologist}, the development of deep learning systems for chest radiography meets a natural demand. The task of chest radiography is well-suited to supervised learning, since copious amounts of chest radiographs and their associated diagnosis labels are being made available as training data for deep learning models.

\begin{figure}[t!]
\begin{center}
\includegraphics[width=1.0\linewidth]{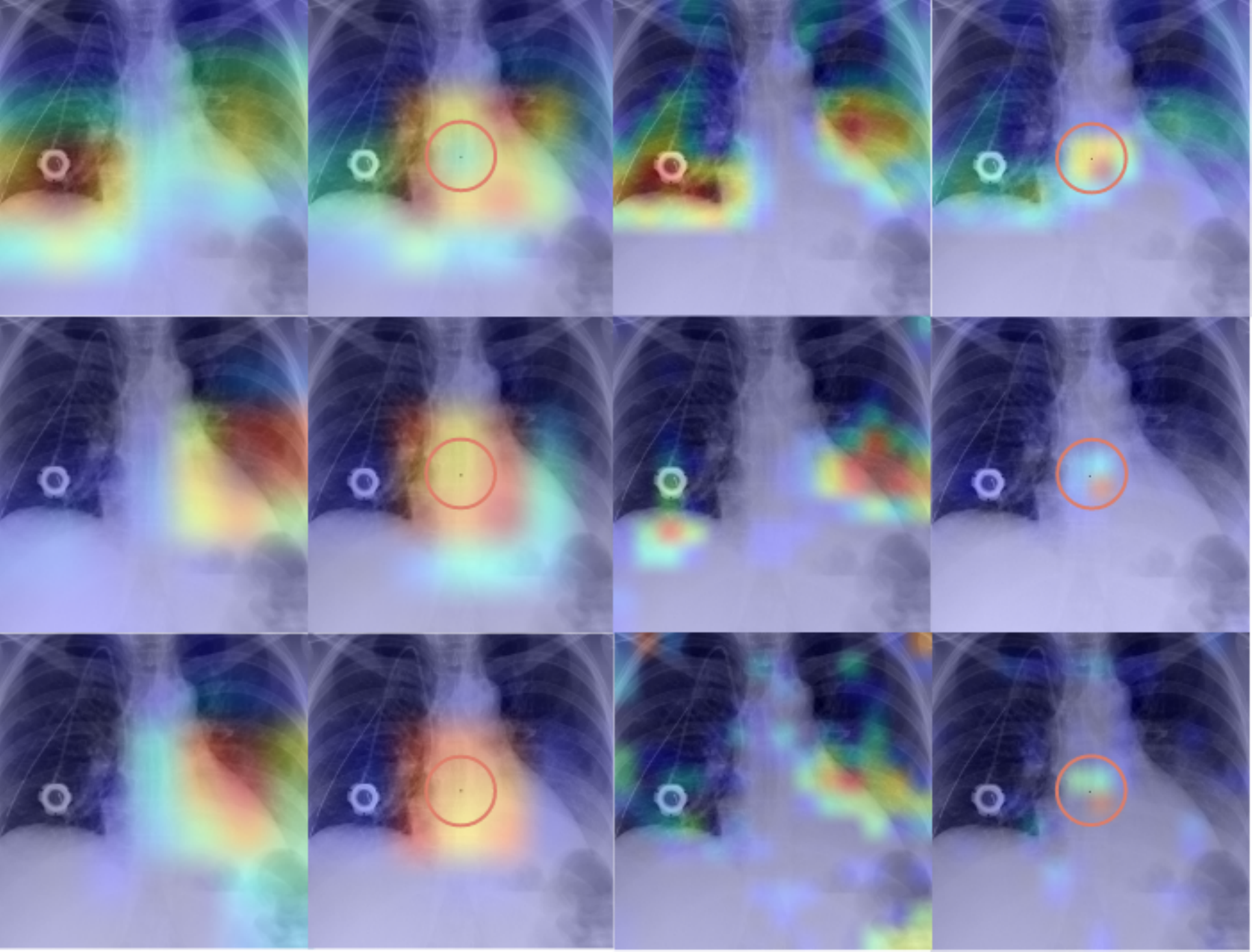}
\end{center}
   \caption{The rows represent epochs 1, 4 and 12 (top to bottom). \textbf{1\textsuperscript{st} Column:} We present Grad-CAM saliency maps with respect to last convolutional layer for a non-infected image.  \textbf{2\textsuperscript{nd} Column:} The maps are taken with respect to the same layer, but for an infected version of the same image.  \textbf{3\textsuperscript{nd} Column:} Moving to a middle convolutional layer, we obtain saliency maps for the non-infected image. \textbf{4\textsuperscript{th} Column:} Finally, we take maps with respect to middle convolutional layer for the infected image. The red circle indicates the location of the trigger.}
\label{fig: explainability}
\end{figure}

Chest radiograph data has been compiled into benchmark datasets for a range of diseases \cite{wang2017chestx,johnson2019mimic}, including the COVID-19 most recently \cite{wang2020covid}. Deep learning models have exhibited expert-level ability in these benchmark datasets \cite{rajpurkar2018deep,seyyed2020chexclusion}. However, it has been recently shown that even the best-performing deep learning models may be susceptible to adversarial attacks \cite{finlayson2019adversarial,han2020deep}. \citet{finlayson2018adversarial} present the argument that the high costs in the medical industry, combined with basis of pharmaceutical device and drug approvals on disease prevalence may motivate manipulation of disease detection systems.

The main contributions of this paper are as follows: 
\begin{enumerate}
    \item We explore backdoor attacks in-depth for medical imaging by focusing on chest radiography in the context of multi-label classification. Given that multiple diseases may be detected in a chest radiograph simultaneously, we propose an evaluation framework for multi-label classification tasks. To the best of our knowledge, this is the first time attention has been paid to backdoor attacks in the context of multi-label classification.
    \item We show how the trigger manifests in both low-level and high-level features learned by the model through Gradient-weighted Class Activation Mapping (Grad-CAM), a weakly-supervised explainability technique \cite{selvaraju2017grad}. By showing how explainability can be used to identify the presence of a backdoor, we emphasize the role of explainability in investigating model robustness. 
\end{enumerate}

\section{Related Work} 
Earlier defense mechanisms against backdoor attacks often assumed access to the triggers or infected samples \cite{tran2018spectral}, or found distinguishable properties in spectral signatures in activation patterns \cite{chen2018detecting}. It has been shown that these techniques can be circumvented by slightly different trigger design \cite{tan2019bypassing, bagdasaryan2020blind, liao2018backdoor}. In a realistic scenario, a defender would at most have access to the model and a small validation set.  Several solutions that do not depend on access to infected samples focus on analyzing the model in which the backdoor has been injected \cite{CCS_ABS, wang2019neural, CCS_ABS, nips_generative18, liu2018fine}. However, these solutions are restricted by assumptions on the trigger size or by the number of neurons needed to encode the trigger. Several attacks on the state-of-the-art defenses highlight that they were designed only for specific triggers and lack generalizability. Most of the defenses fail if more than one label is infected, which again points to a highly constrained defense scenario. Moreover, all of the defenses are tailored for multi-class problems, and therefore cannot be extended to our multi-label backdoor attack.

\section{Methodology}

\begin{figure}[ht]
\begin{center}
\includegraphics[width=1.0\linewidth]{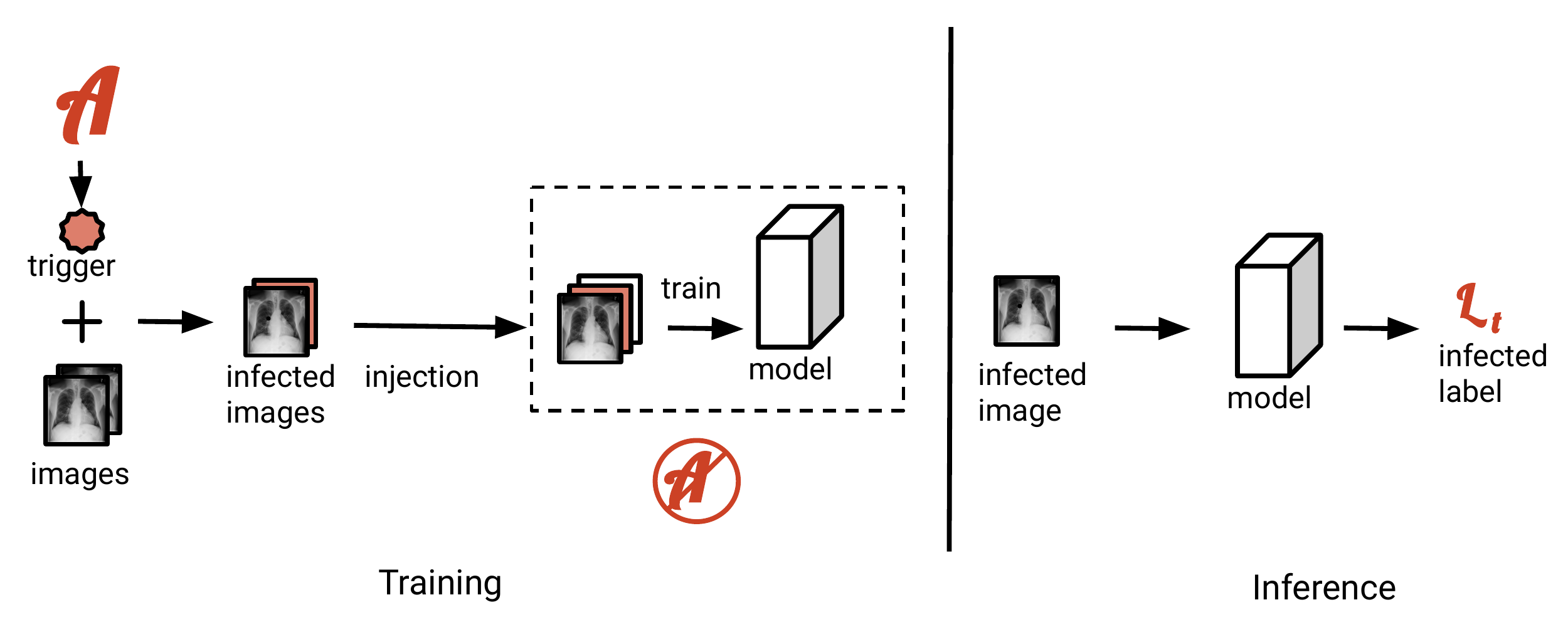}
\end{center}
   \caption{Threat model schematic. Prior to \textbf{training}, the attacker $\mathcal{A}$ inserts a set of images containing backdoor triggers, and their corresponding labels, into the training database. The attacker may not have the benefit of observing or altering the training procedure. During \textbf{inference}, any image containing the trigger is classified to the infected label.
   }
\label{overallmethod}
\end{figure}

\subsection{Threat model}
It is estimated that 3 to 15 percent of the \$3.3 trillion annual healthcare spending in the United States may be attributed to fraud \cite{rudman2009healthcare}. Incentives to manipulate medical imaging systems are present among larger companies seeking pharmaceutical or device approvals \cite{kalb1999health, pien2005using}, as well as insurers or individual practitioners who seek higher compensation or reimbursements based on disease diagnoses \cite{ornstein2014top,reynolds2005metabolic,kesselheim2005overbilling,wynia2000physician}.
 
In our threat model, the \textit{attacker} leverages maliciously  on their special access to the machine learning dataset. They  insert images with backdoor triggers into the training dataset, prior to training, and need not necessarily need to be involved in the training procedure, as shown in Figure \ref{overallmethod}. For instance, they do not need to know how many epochs training will last for, what the hyperparameters of the network will be, or what preprocessing steps may be applied, except for what they may deduce from the nature of the training set, $\mathcal{D}\textsubscript{train}$. The \textit{user} will uknowingly trust the predictions of the infected model, denoted as $\mathcal{M}^\prime$. $\mathcal{M}^\prime$ will be trusted if it achieves some minimum performance \emph{a*} on an independent test set using some metric $\mathcal{A}$. The user does not know that $\mathcal{D}\textsubscript{train}$ has been altered and would only require that $\mathcal{A} (
{\mathcal{M}^\prime}, \mathcal{D}\textsubscript{test}) \geq \emph{a*} $. 
 
\subsection{Attack Formalization}
A backdoor trigger may be applied to a \textit{clean} image $x$ by means of some function $ p(x, r, m) $,
 $$ x^\prime =  p(x, r, m) = x \bullet (1-m) + r\bullet m $$
 
\noindent
 to obtain the infected image $x^\prime$, where $r$ represents the trigger, $m$ denotes the trigger mask that takes a value of 1 at the trigger location and 0 elsewhere, and $\bullet$ is the element-wise product. 
 
We make  distinction, between the true label of a backdoor and the infected label of ${x^\prime}$. The true label of ${x^\prime}$, denoted as $\mathcal{T}(x^\prime)$, is the ground truth set of binary labels associated with ${x}$ for the classes in $\mathcal{D}_{train}$. The infected label of ${x^\prime}$, denoted as $\mathcal{I}$, is the set of output binary labels that the attacker inserts into $\mathcal{D}_{train}$ to be associated with the infected image ${x^\prime}$. We note that $|\mathcal{I}|$ represents the total number of possible disease classes in the dataset and $|\mathcal{I}|=|\mathcal{T}|$. The attacker desires that any image ${x^\prime}$ containing a backdoor trigger will be classified as having a particular target class $t$ with high probability. Therefore, $t$ is set to 1 within $\mathcal{I}$, while all other classes are set to 0.

\subsection{Evaluation Metrics}
To evaluate the success of the backdoor attack, we propose several evaluation metrics in the context of a multi-label classification task.
 
\textbf{Attack success rate:} We define Attack Success Rate (ASR) as the proportion of infected images where the target class $t$ prediction exceeds a minimum confidence $p$ (i.e., $\mathcal{M}^\prime_t\geq p$), within the subset of infected images where the target class $t$ was absent in the true label (i.e., $\mathcal{T}_t=0$). More formally,
$$ ASR = \frac{\sum_{x^\prime : \mathcal{M}^\prime(x^\prime)_{t} \geq p } 1 }{ \sum_{x^\prime:\mathcal{T}(x^\prime)_t = 0} 1 }. $$ 

$p$ is the minimum probability that must be predicted by the model for target class $t$ in order to consider the backdoor attack successful. 

While our formulation of ASR lends insight into how well a backdoor injection succeeds relative to the target class, it does not assess how the backdoor affects classification of all possible classes. It is for this reason that we evaluate the infected model using three variations of the micro-average Area Under the Receiver Operating Characteristics (AUROC) curve metric across all possible labels:

\begin{itemize}
  \item \textbf{AUROC-NN}, where `NN' stands for `Normal image, Normal label'. AUROC-NN measures how well the model predicts the \textit{true labels} of clean (normal) images, after the model is injected with a backdoor during training.  
  
  \item \textbf{AUROC-TT}, where `TT' stands for `Triggered image, Triggered label'. AUROC-TT measures how well the model learns to predict the \textit{infected labels} of infected (triggered) images. However, AUROC-TT does not indicate how well the model misclassifies the infected images, away from the true labels.  
  
  \item \textbf{AUROC-TN}, where `TN' stands for `Triggered image, Normal label'. AUROC-TN measures how well the model misclassifies the infected images away from the true (normal) labels prior to infecting the image. Unlike ASR and AUROC-TT, a lower AUROC-TN score implies a better backdoor performance, and a higher AUROC-TN score implies a poorer backdoor performance. 
 
\end{itemize}
 
To understand the worst and best case performance of the backdoor, we report the minimum and maximum values of each metric over model training epochs.

\section{Experiments}

\subsubsection{Dataset:} We explored and evaluated the impact of backdoor attacks on a publicly available chest radiograph dataset that has been prominent in deep learning research \cite{singh2018deep,zhang2020secret}. The NIH Chestx-ray8 dataset is a Health Insurance Portability and Accountability Act (HIPAA)-compliant dataset that contains 112,120 chest radiographs collected from 30,805 patients \cite{wang2017chestx}.  The \textit{true label} of each image is a binary vector indicating the presence or absence of 14 different diseases, where 1 indicates that a disease is present.

\subsubsection{Backdoor Attack:}
 
We injected the backdoor into our DenseNet-121 model, training it on the infected training set four times with four different random seeds. During evaluation, we reported the mean and standard deviation of the metrics computed using the four trained models. We calculated ASR for two thresholds of $p$, 0.6 and 0.9, to prioritise precision and recall, respectively. We make our code, written in Keras 2.3.1 \cite{chollet2018keras} and Tensorflow 2.0.0 \cite{abadi2016tensorflow}, available.

\subsection{Effect of trigger}

\begin{figure*}[!h]
\begin{center}
\begin{subfigure}{.45\textwidth}
  \centering
  \includegraphics[width=0.82\linewidth]{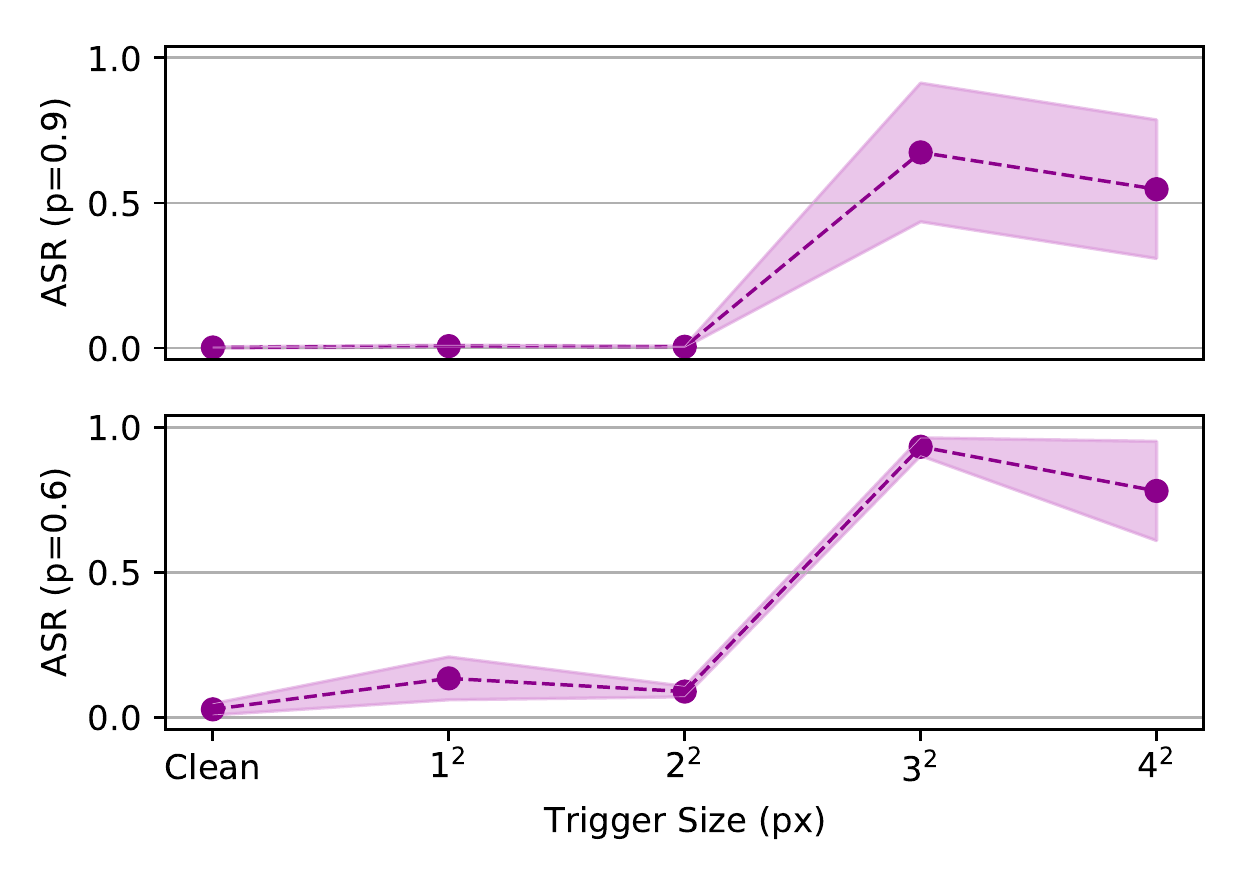}  
  \caption{} 
  \label{fig:asrs}
\end{subfigure}
\begin{subfigure}{.45\textwidth}
  \centering
  \includegraphics[width=1.0\linewidth]{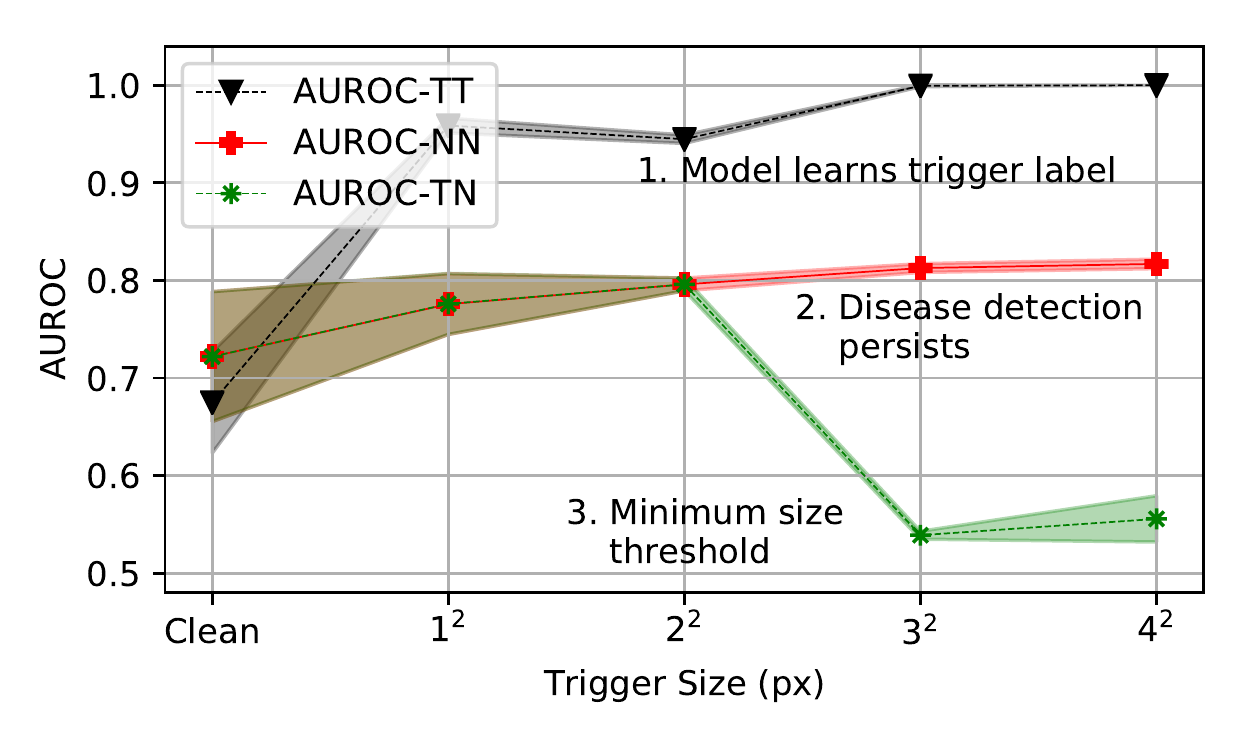}  
  \caption{} 
  \label{fig:aurocs}
\end{subfigure}
\caption{(a) We demonstrate the effect of increasing the trigger size in training on the backdoor attack success rate in inference, using our two probability thresholds of choice. The attacks become more successful as the trigger size grows beyond 2 squared pixels. (b) Variations in the AUROC are measured against trigger size.  Notice that AUROC-TN drops noticeably as the trigger size increases beyond 2 squared pixels. AUROC-TT approaches value of 1. However, the AUROC-NN values remain level, indicating that performance of the neural network on a clean dataset appears unaffected to the unsuspecting \textit{user}. In both (a) and (b), center lines represent mean values, while the surrounding regions represent standard deviations, to scale.}
\label{fig:asrsaurocs}
\end{center}
\end{figure*}

Our backdoor trigger, designed to suit chest the grayscale of radiograph datasets, consists solely of black pixels.  We ran a number of experiments to study trigger behaviour, adding triggers to 40\% of training images without replacement.
 
We ran experiments to verify whether the performance of the backdoor attack is invariant to the location of the trigger. Specifically, in comparison to placing the trigger in a fixed location (center of image), trained the model to recognize the backdoor in a randomly chosen location within the image boundary. We present our results in Table \ref{table: random_fixed}. The ASR values remain high at $\geq$0.963  $(p=0.6)$, AUROC-TT values are near 1, and AUROC-TN values are low ($\leq$0.704), showing the backdoor attack is effective regardless of its location on the image.
    
We also experimented with black backdoor triggers of size varying between 1$\times$1 (single pixel) and 4$\times$4 pixels to understand how trigger size affects backdoor performance. Results in Table \ref{table: triggersize_max} suggests that even a single-pixel backdoor trigger can attain high maximum ASR values. AUROC-NN only  vary  slightly with trigger size, meaning it will be hard for a user to detect model infection. The minimum values of the evaluation metrics across all epochs reveal the attacker's worst case performance of their backdoor, visualized in Figure \ref{fig:asrsaurocs}. Minimum ASR, increases sharply at the 3$\times$3 and 4$\times$4 triggers, while AUROC-TN drops lower at those values. AUROC-TT, which is less sensitive to false negatives than AUROC-TN, also increases slightly. Overall, the backdoor attack is more successful as the trigger size increases.  

To understand how much an attacker can use their backdoor during inference time without raising suspicions, we analyzed the performance of the infected model on a test set containing clean and infected images. Table \ref{table: inference_rate} summarizes the results. We find that by keeping $\varepsilon$ small, the drop in the AUROC due to infected images remains comparable to the AUROC on a set of clean images. For $\varepsilon = 0.001$, the maximum AUROC achieved is 0.850. However, as $\varepsilon$ increases, $\varepsilon\geq 0.1$, the AUROC drops below 0.800. The minimum AUROC performance results follow a similar trend.

\begin{table}[t]
\normalsize
\centering
\caption{The maximum scores over all epochs for various trigger sizes are shown. We observe that unlike other metrics, the maximum AUROC-TN score indicates the worst-case performance, since a higher value of AUROC-TN indicates a lower responsiveness to the presence of a backdoor trigger. We include results for the clean data as a control to demonstrate the effectiveness of the backdoor triggers. }
\resizebox{1.0\linewidth}{!}{
\begin{tabular}{||l||ccccc||}
\hline
Trigger Size & Clean  & $1^2 px$ & $2^2 px$ & $3^2 px$ & $4^2 px$ \\ 
\hline\hline
 ASR($p=0.6$) & $0.574\pm0.269$  & $1.000\pm0.000$  & $1.000\pm0.000$  & $0.999\pm0.000$   & $1.000\pm0.000$  \\
 \hline
 ASR($p=0.9$) & $0.268\pm0.194$  & $1.000\pm0.000$  & $1.000\pm0.000$ & $0.998\pm0.001$  & $0.997\pm0.003$ \\
 \hline
 AUROC-TT & $0.852\pm0.046$ & $1.000\pm0.000$ & $1.000\pm0.000$  &  $1.000\pm0.000$  & $1.000\pm0.000$  \\ 
 \hline
 AUROC-NN & $0.846\pm0.045$  & $0.824\pm0.002$ & $0.824\pm0.001$ & $0.851\pm0.001$ & $0.850\pm0.004$  \\ 
 \hline
 AUROC-TN  & $0.848\pm0.002$ & $0.824\pm0.002$ & $0.824\pm0.001$ & $0.704\pm0.038$ & $0.727\pm0.018$ \\
\hline
\end{tabular}}
\label{table: triggersize_max}
\end{table}

\begin{table}[!h]
\small
\centering
\caption{We obtain results for when trigger is fixed in the center of infected images, compared to performance given a random trigger location. The results demonstrate that the backdoor trigger is learned by the neural network notwithstanding its spatial localization.}
\resizebox{1.0\linewidth}{!}{
\begin{tabular}{||l||cc|cc||} 
\hline 
 & \multicolumn{2}{c|}{Fixed} &\multicolumn{2}{c||}{Random} \\
\cline{2-5}
 & Minimum & Maximum & Minimum & Maximum \\ 
 \hline\hline
 ASR($p=0.6$) & $0.994\pm0.031$ & $0.999\pm0.000$ & $0.963\pm0.028$ & $0.994\pm0.002$  \\
 \hline
 ASR($p=0.9$) & $0.674\pm0.239$ & $0.998 \pm0.001$ & $0.902\pm0.079$ & $0.991\pm0.001$  \\
 \hline
 AUROC-TT & $1.000\pm0.000$ & $1.000\pm0.000$ & $0.998\pm0.001$ & $1.000\pm0.000$\\ 
 \hline
 AUROC-NN & $0.813\pm0.004$ & $0.851\pm0.001$ & $0.815\pm0.007$  & $0.848\pm0.005$  \\ 
 \hline
 AUROC-TN  & $0.539\pm0.004$ & $0.704\pm0.038$ & $0.579\pm0.027$ & $0.696\pm0.008$  \\ 
\hline
\end{tabular}}
\label{table: random_fixed}
\end{table}

\begin{table}[!h]
\small
\centering
\caption{ $\varepsilon$ represents the proportion of infected images in the test set containing clean and infected images. The results suggest that the attacker may successfully use their backdoor attack during inference without drawing significant attention from the user by maintaining a low proportion of infected images during inference time.
}
\resizebox{1.0\linewidth}{!}{
\begin{tabular}{||l||cc|cc||} 
\hline 
  & \multicolumn{4}{c||}{\textbf{$\varepsilon$}} \\
\cline{2-5}
 & 0.001 & 0.01 & 0.1 & 0.5 \\ 
 \hline\hline
 Minimum & $0.809\pm0.005$ & $0.804\pm0.005$ & $0.793\pm0.001$  & $0.604\pm0.002$  \\
 \hline
 Maximum & $0.852\pm0.001$ & $0.847\pm0.001$ & $0.761\pm0.004$  & $0.662\pm0.006$  \\
\hline
\end{tabular}}
\label{table: inference_rate}
\end{table}

\subsection{Explainability can localize backdoor triggers}

Explainability techniques are commonly used in medical imaging applications \cite{holzinger2019causability}. We therefore examined the role of explainability in the context of backdoor attacks using Gradient Class Activation Mappings (Grad-CAM) \cite{selvaraju2017grad}. Grad-CAM calculates the derivative of activations with respect to the last convolutional layer of the neural network to compute saliency maps based on high-level features, where more important regions are indicated by red and less important regions are indicated by blue. We take this one step further by applying Grad-CAM to a middle layer, to examine low-level features. DenseNet-121 contains 287 layers, and we chose the 207th layer as the middle layer, since the  dimensions of the feature maps in this layer match those of the final layer.

Figure \ref{fig: explainability} shows the computed saliency maps for the clean and infected versions of an example image, at epochs 1, 4, and 12 from top to bottom. First, we note that the inclusion of a backdoor trigger causes a change in the heatmap, when comparing the changes in the saliency maps with respect to the final convolutional layer (columns 1 and 2), and the middle convolutional layer (columns 3 and 4) at each epoch. In particular, we observe that in a clean image (column 1), the heatmap focuses on the center of the patient's lungs. However, in column 2, the heatmap shifts towards the sternum region of the radiograph, where the backdoor trigger is located, across all epochs.

 As can be seen in columns 3 and 4, the middle layer shows more fine-grained backdoor trigger localization. While there are some saliency artifacts in column 3, we notice that the localization heatmap shift to the center of the image where the trigger is located in column 4. This is more noticeable at epochs 1 and 4, where there is less overfitting. The increased localization in the middle of the network is understandable since the backdoor trigger pixels can be considered as low-level features, and thus may be better detected in the earlier layers of the network. This suggests that explainability can play a complementary role with robustness, since Grad-CAM shows differences between the saliency maps of clean and infected images, and can help radiologists in questioning model predictions when the saliency maps and predictions seem unreasonable. One limitation of the work is that we did not investigate instances where the location of the trigger could correspond rightfully so with expected clinical interpretation. This requires clinical domain expertise.

\section{Conclusion}
In this work, we propose a framework for evaluating backdoor effectiveness in the multi-label setting. Focusing on the medical imaging task of chest radiography, we show how explainability is a valuable tool in backdoor trigger detection. Future work should investigate the impact of backdoors and explainability on other medical imaging tasks, and the design of suitable defense mechanisms.

 
\bibliography{references}
\clearpage

\newpage

\end{document}


\maketitle
 \vspace{-10cm}

\begin{table*}[h!]

\caption{To check that our backdoor is not dependent on the choice of target class, we vary the target class of our backdoor experiments to cover all classes in our dataset, using our 3$\times$3 pixel trigger. The maximum and minimum values attained per class are presented in (a) and (c) respectively. We observe that the backdoor is compelling for most classes. We also perform experiments to gain insight into the ideal proportion of infected images to be injected into the training set $\mathcal{D}\textsubscript{train}$. The maximum and minimum values attained for each given proportion are presented in (b) and (d) respectively. A smooth upward trend in ASR, and downward trend in AUROC-TN is observed. The greater the percentage of infected images in training, the more strongly the neural network learns the trigger feature. We visualise this effect in our main paper in figure \ref{fig: trigger_perc}.  }

\begin{subtable}[t]{1.0\textwidth}

\small
\caption{}
\begin{center}
\resizebox{0.6 \linewidth}{!}{
\begin{tabular}{||l||cc|ccc||}
\hline
Class & ASR(p=0.6) & ASR(p=0.9) & AUROC-TT & AUROC-NN & AUROC-TN \\
\hline\hline
 Atelectasis      & $0.999\pm0.000$  & $0.998\pm0.001$  & $1.000\pm0.000$ & $0.851\pm0.001$  & $0.704\pm0.038$ \\
 \hline
 Cardiomegaly      & $0.991\pm0.001$ & $0.988\pm0.000$ & $1.000\pm0.000$ & $0.851\pm0.004$ & $0.650 \pm0.021$ \\
 \hline
 Effusion          & $0.993\pm0.001$ & $0.989\pm0.002$ & $1.000\pm0.000$ & $0.849\pm0.002$ & $0.704\pm0.019$ \\
 \hline
 Infiltration      & $0.995\pm0.002$ & $0.991\pm0.001$ & $1.000\pm0.000$ & $0.851\pm0.002$ & $0.749\pm0.026$ \\ 
 \hline
 Mass              & $0.990\pm0.001$ & $0.985\pm0.002$ & $1.000\pm0.000$ & $0.850\pm0.003$ & $0.676\pm0.031$ \\
 \hline
 \hline
 Nodule            & $0.070\pm0.034$ & $0.020\pm0.014$ & $1.000\pm0.000$ & $0.847\pm0.004$ & $0.847\pm0.004$ \\
 \hline
 Pneumonia         & $0.990\pm0.001$ & $0.987\pm0.003$ & $1.000\pm0.000$ & $0.850\pm0.003$ & $0.654\pm0.021$ \\
 \hline
 Pneumothorax      & $0.993\pm0.001$ & $0.990\pm0.001$ & $1.000\pm0.000$ & $0.850\pm0.003$ & $0.686\pm0.014$ \\
 \hline
 Consolidation     & $0.993\pm0.002$ & $0.989\pm0.004$ & $1.000\pm0.000$ & $0.849\pm0.001$ & $0.669\pm0.021$ \\
 \hline
 Edema             & $0.993\pm0.002$ & $0.990\pm0.003$ & $1.000\pm0.000$ & $0.850\pm0.003$ & $0.627\pm0.017$ \\
 \hline
 Emphysema         & $0.992\pm0.002$ & $0.988\pm0.002$ & $1.000\pm0.000$ & $0.850\pm0.002$ & $0.664\pm0.020$ \\
 \hline
 Fibrosis          & $0.990\pm0.002$ & $0.987\pm0.003$ & $1.000\pm0.000$ & $0.850\pm0.003$ & $0.647\pm0.026$ \\
 \hline
 Pleural Thickening& $0.990\pm0.001$ & $0.986\pm0.003$ & $1.000\pm0.000$ & $0.849\pm0.003$ & $0.651\pm0.014$ \\
 \hline
 Hernia            & $0.991\pm0.001$ & $0.988\pm0.002$ & $1.000\pm0.000$ & $0.850\pm0.003$ & $0.651\pm0.025$ \\
\hline
\end{tabular}}
\end{center}
\label{table: classcompsmax}
\end{subtable} 
\vspace{-1em}

\begin{subtable}[t]{1.0\textwidth}

\small
\caption{}
\begin{center}
\resizebox{0.7\linewidth}{!}{
\begin{tabular}{||l||c|c|c|c|c|c|c||}
\hline
Class & 1 & 5 & 10  & 20  & 40  & 100 & $\mathcal{D}_{trig}$ \\
\hline\hline
 ASR(p=0.6) &$0.913\pm0.048$& $0.973\pm0.002$ & $0.972\pm0.021$ & $0.988\pm0.002$ & $0.999\pm0.000$ & $0.997\pm0.000$ & $1.000\pm0.000$  \\
 \hline
 ASR(p=0.9) & $0.841\pm0.053$ & $0.943\pm0.017$ & $0.944\pm0.034$ & $0.982\pm0.001$ & $0.998\pm0.001$ & $0.994\pm0.001$ & $1.000\pm0.000$ \\
 \hline
 AUROC-TT   & $0.998\pm0.001$ & $1.000\pm0.000$ & $1.000\pm0.000$ & $1.000\pm0.000$ & $1.000\pm0.000$ & $1.000\pm0.000$ & $1.000\pm0.000$ \\
 \hline
 AUROC-NN   & $0.843\pm0.006$ & $0.851\pm0.006$ & $0.851\pm0.004$ & $0.851\pm0.003$ & $0.851\pm0.001$ & $0.876\pm0.042$ & $0.536\pm0.003$ \\ 
 \hline
 AUROC-TN   & $0.821\pm0.015$ & $0.776\pm0.016$ & $0.774\pm0.026$ & $0.742\pm0.009$ & $0.704\pm0.038$ & $0.638\pm0.035$ & $0.535\pm0.004$ \\
 \hline
\end{tabular}}
\end{center}
\label{table: perccompsmax}
\end{subtable} 
\vspace{-1em}

\begin{subtable}[t]{1.0\textwidth}

\small
\caption{}
\begin{center}
\resizebox{0.6\linewidth}{!}{
\begin{tabular}{||l||cc|ccc||}
\hline
Class & ASR(p=00.6) & ASR(p=0.9) & AUROC-TT & AUROC-NN & AUROC-TN \\
\hline\hline
 Atelectasis      & $0.994\pm0.031$  & $0.674\pm0.239$  & $1.000\pm0.000$  & $0.813\pm0.004$   & $0.539\pm0.004$  \\
 \hline
 Cardiomegaly      & $0.962\pm0.008$ & $0.926\pm0.022$ & $1.000 \pm0.000$ & $0.810\pm0.008$ & $0.535\pm0.013$ \\
 \hline
 Effusion          & $0.973\pm0.012$ & $0.926\pm0.032$ & $1.000 \pm0.000$ & $0.813\pm0.004$ & $0.582\pm0.013$ \\
 \hline
 Infiltration      & $0.963\pm0.030$ & $0.757\pm0.273$ & $1.000 \pm0.001$ & $0.814\pm0.011$ & $0.656\pm0.014$ \\ 
 \hline
 Mass              & $0.866\pm0.065$ & $0.960\pm0.006$ & $1.000 \pm0.000$ & $0.805\pm0.008$ & $0.552\pm0.013$ \\
 \hline
 \hline
 Nodule            & $0.009\pm0.012$ & $0.001\pm0.001$ & $0.999\pm0.001$ & $0.810\pm0.004$ & $0.810\pm0.004$ \\
 \hline
 Pneumonia         & $0.927\pm0.056$ & $0.968\pm0.015$ & $1.000 \pm0.000$ & $0.811\pm0.004$ & $0.514\pm0.017$ \\
 \hline
 Pneumothorax      & $0.975\pm0.007$ & $0.936\pm0.023$ & $1.000 \pm0.000$ & $0.812\pm0.008$ & $0.553\pm0.017$ \\
 \hline
 Consolidation     & $0.922\pm0.055$ & $0.808\pm0.158$ & $0.999\pm0.001$ & $0.798\pm0.020 $ & $0.524\pm0.011$ \\
 \hline
 Edema             & $0.977\pm0.004$ & $0.951\pm0.007$ & $1.000 \pm0.000$ & $0.806\pm0.010 $ & $0.505\pm0.021$ \\
 \hline
 Emphysema         & $0.964\pm0.007$ & $0.889\pm0.025$ & $1.000 \pm0.000$ & $0.802\pm0.009$ & $0.532\pm0.011$ \\
 \hline
 Fibrosis          & $0.954\pm0.013$ & $0.884\pm0.041$ & $0.999\pm0.000$ & $0.808\pm0.003$ & $0.516\pm0.017$ \\
 \hline
 Pleural Thickening& $0.956\pm0.016$ & $0.878\pm0.044$ & $1.000 \pm0.001$ & $0.807\pm0.011$ & $0.522\pm0.017$ \\
 \hline
 Hernia            & $0.941\pm0.047$ & $0.871\pm0.094$ & $1.000 \pm0.000$ & $0.807\pm0.013$ & $0.522\pm0.012$ \\
\hline
\end{tabular}}
\end{center}
\label{table: classcompsmin}
\end{subtable} 

\vspace{1em}

\begin{subtable}{1.0\textwidth}
\small
\caption{}
\begin{center}
\resizebox{0.7\linewidth}{!}{
\begin{tabular}{||l||c|c|c|c|c|c|c||}
\hline
Class & 1 & 5 & 10 & 20 & 40 & 100 & $\mathcal{D}_{trig}$ \\
\hline\hline
 ASR(p=0.6) & $0.343\pm0.232$& $0.705\pm0.103$ & $0.599\pm0.195$ & $0.939\pm0.021$ & $0.994\pm0.031$ & $0.986\pm0.005$ & $0.000\pm0.000 $  \\
 \hline
 ASR(p=0.9) & $0.092\pm0.134$ & $0.278\pm0.191$ & $0.262\pm0.214$ & $0.750 \pm0.085$ & $0.674\pm0.239$ & $0.957\pm0.017$ & $0.000  \pm0.000 $ \\
 \hline
 AUROC-TT   & $0.844\pm0.059$ & $0.968\pm0.041$ & $0.965\pm0.036$ & $0.999\pm0.001$ & $0.000\pm0.000$ & $0.000\pm0.000$ & $0.000\pm0.000$ \\
 \hline
 AUROC-NN   & $0.761\pm0.044$ & $0.796\pm0.010 $ & $0.803\pm0.011$ & $0.816\pm0.006$ & $0.813\pm0.004$ & $0.812\pm0.003$ & $0.486\pm0.038$  \\
 \hline
 AUROC-TN   & $0.588\pm0.030$ & $0.572\pm0.019$ & $0.620 \pm0.020$ & $0.591\pm0.032$ & $0.539\pm0.004$ & $0.547\pm0.007$ & $0.486\pm0.039$ \\
\hline
\end{tabular}}
\end{center}
\label{table: perccompsmin}

\end{subtable} 

\end{table*}


\newpage